\documentclass[10pt]{iopart}
\usepackage{iopams}  
\usepackage{graphicx}
\usepackage{xcolor}
\usepackage{cite}
\expandafter\let\csname equation*\endcsname\relax
\expandafter\let\csname endequation*\endcsname\relax
\usepackage{amsmath}
\usepackage{tablefootnote}
\begin{document}

\title[]{Breaking the wire: the impact of critical length on melting pathways in silver nanowires}
\author{K M Ridings$^{1,2}$, E E L Vaka'uta$^2$, S M Croot$^2$}
\address{$^1$MacDiarmid Institute for Advanced Materials and Nanotechnology,} 
\address{$^2$Department of Physics, University of Auckland, Private Bag 92019, Auckland, New Zealand.}

\ead{k.ridings@auckland.ac.nz}
\vspace{10pt}
\begin{indented}
\item[]November 2024
\end{indented}

\begin{abstract}
We explore the melting mechanisms of silver nanowires through molecular dynamics simulations and theoretical modeling, where we observe that two distinct mechanisms or pathways emerge that dictate how the solid-liquid interface melts during the phase transition. For wires longer than a critical length ($L>L_{\textrm{crit}}$), an Arrhenius-type diffusion model successfully predicts the solid-liquid interface velocity, highlighting diffusion-driven melting pathways. In contrast, wires shorter than the critical length ($L\leq L_{\textrm{crit}}$) exhibit unique behaviors driven by non-equilibrium effects, including rapid overheating of the solid core, stabilization of the solid-liquid interface, and the pronounced impact of higher energy densities. These mechanisms lead to accelerated melting and distinct phase transition dynamics. Our findings reveal how geometry and nanoscale effects critically shape melting behavior, offering insights for the design and stability of nanostructures in advanced applications.
\end{abstract}
%
%
%
\maketitle
%
\ioptwocol
\section{Introduction}
It is well known that nanostructured objects have lower melting points than their bulk counterparts \cite{wronski1967size, coombes1972melting, di1995maximum, ridings2019surface}. Due to the lower surface-to-volume ratio in nanostructures, they have a lower stability with respect to their molten phase \cite{bachels2000melting, shin2007size}. The presence of Plateau-Rayleigh instabilities in nanostructures can break apart geometries such as rings, thin films, and nanowires into chains of nanospheres have been observed \cite{toimil2004fragmentation, dutta2014silico,xu2018situ, nguyen2012competition, allaire2021role}. The lowering of nanostructure stability with respect to their melts combined with Plateau-Rayleigh type instabilities describes how nanowire or nanofilament stability, where longer nanowires or filaments appear to be more unstable, and can even melt at slightly lower temperatures than their shorter counterparts \cite{wu2022molecular, ridings2022nanowire}. Understanding the stability limits of nanowires or nanofilaments has important applications in things such as neuromorphic nanoparticles networks \cite{wu2022molecular, wu2023neuromorphic}, where understanding conditions leading to nanowire or nanofilament breakage are crucial to design functionality. Melting pathways in Ag nanoclusters have been observed where melting and non-melting modes appear and are sensitive to system size \cite{liang2017size}. Additionally, it has been observed that the melting pathway in nanowires depends on their aspect ratio, where two distinct modes appear \cite{ridings2019surface, ridings2022nanowire}. With this in mind, we want to ask why does the emergence of these distinct modes arise in the first place, and what mechanisms dictate this behaviour? One way researchers study the melting or crystallization process in different materials is to model the growth of the solid-liquid interface \cite{wilson1900xx, Frenkel1932, broughton1982crystallization, coriell1982relative}. Understanding the growth (or decay) of a crystal is important so researchers can competently control the growth of nanostructures. When constructing models on crystallization of melting, using the interface velocity of a material is a common method \cite{wu2021crystal, wu2021crystal_2, sun2018mechanism, mazhukin2020atomistic}. These models are simple, and take the general form:
\vspace{0.5cm}
\\
\begin{equation}
    \label{Eq:int_velocity}
    V(T) = k(T)(1- \exp(-\Delta\mu/k_{\text{B}} T)),
\end{equation}

where  $k(T)$ is a kinetic term, $k_{\text{B}}$ is the Boltzmann constant, and $
\Delta \mu$ is the difference in chemical potential between the solid and liquid phases. $\Delta \mu$ can be expressed in terms of the undercooling in a material, i.e. $\Delta \mu \approx \Delta h_{\text{m}} \Delta T / T_{\text{c}}$, where $\Delta h_{\text{m}}$ is the enthalpy of fusion per particle at the melting temperature, $\Delta T$ is the undercooling, and $T_{\text{c}}$ is the bulk melting temperature of the material. Researchers have noted the specific form the kinetic coefficient $k(T)$ takes can tell us what mechanisms are driving the growth (or decay) of a solid-liquid interface \cite{sun2018mechanism}.
\\
In this paper, we explore the mechanisms behind the melting modes observed in metal nanowires as their length is increased, using Ag nanowires in particular. Using a simple interface velocity model, we will examine how well a diffusion-type of model can predict the observed melting pathways wires of increasing lengths take. We then utilise some simple arguments from classical nucleation theory to calculate thermodynamic quantities that are then substituted into the proposed interface velocity model. When required, we make claims for why the model breaks down (particularly in the case for the shortest wires). 
\section{Thermodynamics and Liquid Nucleus Growth Model}
\begin{figure}[htp]
	\centering
{\includegraphics[width=0.50\textwidth]{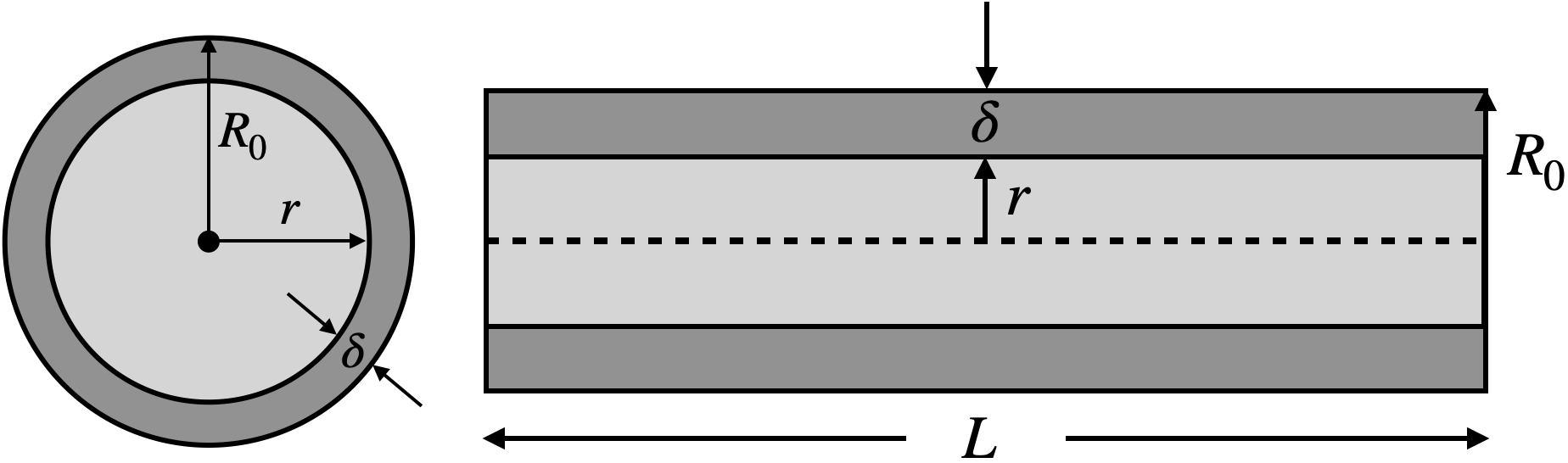}}
    \caption{A top-down view of a surface melted nanowire of radius $R_0$, length $L$, inside solid radius $r$, and liquid nucleus width $\delta = R_0 - r$.}
	\label{Fig:NW_Defns}
\end{figure}

We begin by defining the relevant geometric definitions of a surface-melted nanowire of radius $R_0$, length $L$, inside solid radius $r$, and liquid nucleus width defined as $\delta=R_0-r$, as seen in figure~\ref{Fig:NW_Defns}. 
In this paper and previous work, we see the nanowire aspect ratio, $L/R_0$, plays an important role in the melting pathway a nanowire takes \cite{ridings2019surface, ridings2022nanowire}.
We now turn to using simple arguments from classical nucleation theory and interface velocity models to formulate our ideas \cite{ridings2022nanowire,bai2006calculation, wu2015self,broughton1982crystallization,jackson2002interface,sun2018mechanism,mazhukin2020atomistic,wu2021crystal}. Classical nucleation theory tells us that the Gibbs free energy in a coexisting cylindrically symmetric nanowire of radius $r$ surrounded by its own melt can be expressed as
\begin{equation}
    \label{eq:gibbs}
    \Delta G = - \pi r^2 L \Delta G_{\text{v}} + 2 \pi r L \gamma_{\text{sl}},
\end{equation}

where $\gamma_{\text{sl}}$ is the solid-liquid interfacial energy, $\Delta G_{\text{v}} \approx L_\text{v} \Delta T/T_{\text{c}}$ is the Gibbs free energy difference per unit volume, $L_{\text{v}}$ is the bulk latent heat of melting per unit volume, $r$ is the radius of the inside solid and $L$ is the nanowire length. Using the definition of $\Delta G_{\text{v}}$ we find the radius of the solid that minimizes the Gibbs free energy\
\begin{equation}
    \label{eq:rstar}
    r^* = \frac{\gamma_{\text{sl}} T_{\text{c}}}{L_{\text{v}}\Delta T}.
\end{equation}

Substituting this into $\Delta G$ gives the barrier height that separates the solid and liquid phases
\begin{equation}
    \label{eq:gibbs_star}
    \Delta G^* = \frac{\pi \gamma_{\text{sl}}^2 T_{\text{c}} L}{L_{\text{v}} \Delta T}.
\end{equation}

If we assume that one melting mode is dictated by Arrhenius diffusion \cite{broughton1982crystallization}, then the kinetic coefficient $k(T)$ will take a form
\begin{equation}
    \label{eq:Diff_Arrhenius}
    k(T) = C\, \exp(-Q / k_{\text{B}}T).
\end{equation}

where $Q$ represents an activation energy to diffusion, and we assume $C$ depends on the diffusivity of the material, temperature, and wire aspect ratio. This leads to an expression for the interface velocity as
\begin{equation}
    \label{eq:Vel_Diff_Arrhenius}
    V(T) = C \,\exp(-Q / k_{\text{B}}T)(1-\exp(-\Delta \mu / k_{\text{B}}T)).
\end{equation}

We can use this model to understand the behavior of the solid-liquid interface and how the nanowire aspect ratio influences its dynamic evolution.
\section{Computational Details}
We employ molecular dynamics simulation in LAMMPS to conduct this study \cite{plimpton1995fast}. An EAM potential is used to model interactions between atoms, and gives good results for the melting temperatures \cite{sheng2011highly}. To account for the expansion of the lattice, we take the density of Ag at the melting temperature, calculating the volume per atom as $V_{\text{atom}} = m/\rho$. Then equating this to total volume divided by the total number of atoms in a $4a\times 4a\times 100a$ box (where $a$ is the lattice spacing), we get $m/\rho = V_{\text{Tot}}/N_{\text{box}}$ (where for the defined box a total of 6400 atoms is obtained). Solving for the lattice constant gives $a=4.246$ \AA\cite{wang2013modified}. 
\\
A timestep of 2.5 fs was used, and the temperature controlled with a Langevin thermostat using a damping parameter $\gamma = 1.0$ ps$^{-1}$. Periodic boundary conditions in all directions were used, effectively simulating an infinitely long nanowire. Additionally, this suppressed long-wavelength disturbances that may otherwise break the wire prior to complete melting. This ensured a quick equilibration at each timestep of the simulation. The simulations were initialized at an initial temperature of $T_{\text{i}}$ for 1.0 ns. Then a production phase of 0.40 ns increased the temperature from $T_\text{i}$ to $T_\text{i}+1$ was performed, followed by a subsequent equilibration phase of 0.6 ns at temperature $T_\text{i} +1$. This gave an effective heating rate of 1.0 K/ns. This was to ensure each nanowire was close to equilibrium.
\\
We use wires with an initial radius of $R_0 \approx 34$ \AA\, and the following definitions are listed in table~\ref{tab:NW_defns}. Wires aspect ratios were chosen to ensure that a broad range of lengths were used, with $L_1 \ll L_{\text{crit}}$, $L_2 < L_{\text{crit}}$, $L_3 \simeq L_{\text{crit}}$, and $L_4$ to $L_7 \gg L_{\text{crit}}$, where $L_{\text{crit}} \simeq 2 \pi R_0$.
\begin{table*}[htp]
\caption{\label{tab:NW_defns}A table showing the wire aspect ratios $L/R_0$, lengths $L$, and number of particles $N_{\textrm{p}}$. An initial radius of $R_0 = 34$ \AA\ is used.}
\begin{tabular*}{\textwidth}{@{}l*{2}{@{\extracolsep{0pt plus
12pt}}l}}
\br
Aspect Ratio $L/R_0$&Length L (\AA)& Number of particles $N_{\text{p}}$\\
\mr
2.5  & 84  & 19381  \\
5.0  & 168  & 38281  \\ 
7.5  & 254  & 57181  \\
10.0  & 338  & 76081  \\
15.0  & 508  & 113881  \\
25.0  & 848 & 189481  \\
37.5  & 1272  & 283981  \\
\br
\end{tabular*}
\end{table*}
In order to track the evolution of the solid-liquid interface, an order parameter is needed to classify atoms in either a `solid-like' and `liquid-like' state. As such, we employed a type of Steinhardt parameter and took its average value, which defines a local bond order parameter $\overline{\tilde{q_6}}(i)$ to every atom in the system \cite{steinhardt1983bond, lechner2008accurate,eslami2018local} (where the parameter defined by Eslami et al was used). This particular Steinhardt parameter has a broad bimodal distribution for a system in solid-liquid coexistence, meaning a `solid-like' and `liquid-like' state can be defined. These order parameters use spherical harmonics to that determine how connected a central particle is to its neighbours and take the form:
\begin{equation}
    \label{qlm_1}
    q_{lm}(i) = \frac{1}{N_{\text{b}}(i)}\sum_{j\in N_{{\text{b}}}} Y_{lm}(\theta_{ij},\phi_{ij}),
\end{equation}

where $N_{\text{b}}(i)$ represents the number of neighbours of atom $i$, $j$ is one of the neighbours of atom $i$, $\theta_{ij}$ and $\phi_{ij}$ are the azimuthal and polar angles pointing from particles $i$ to $j$, and $Y_{lm}$ are the spherical harmonics. Then the order parameter is averaged over the number of functions defining the order parameter:
\begin{equation}
    \label{qlm_2}
    q_{l}(i) = \bigg( \frac{4 \pi}{2l + 1}\sum_{m=-l}^{m=l} |q_{lm}(i)|^2\bigg),
\end{equation}

where $l$ is a free integer parameter, and $m$ is an integer that runs from $-l$ to $l$. By averaging the dot products over the $\textbf{q}_l$ vectors, a local order parameter can be defined:
\begin{equation}
    \label{qlm_3}
    \Tilde{q_{l}}(i) = \textbf{q}_l(i)\cdot \textbf{q}_l(j) = \frac{1}{N_{\text{b}}(i)}\sum_{j\in N_{{\text{b}}}} \sum_{m=-l}^{m=+l}\hat{q_{lm}}(i)\hat{q_{lm}}^*(j),
\end{equation}

where 
\begin{equation}
\label{qlm_4}
\hat{q_{lm}}(i) = \frac{q_{lm}(i)}{\big( \frac{4 \pi}{2l + 1}\sum_{m=-l}^{m=l} |q_{lm}(i)|^2\big)}.
\end{equation}

\begin{figure}[htp]
	\centering
{\includegraphics[width=0.50\textwidth]{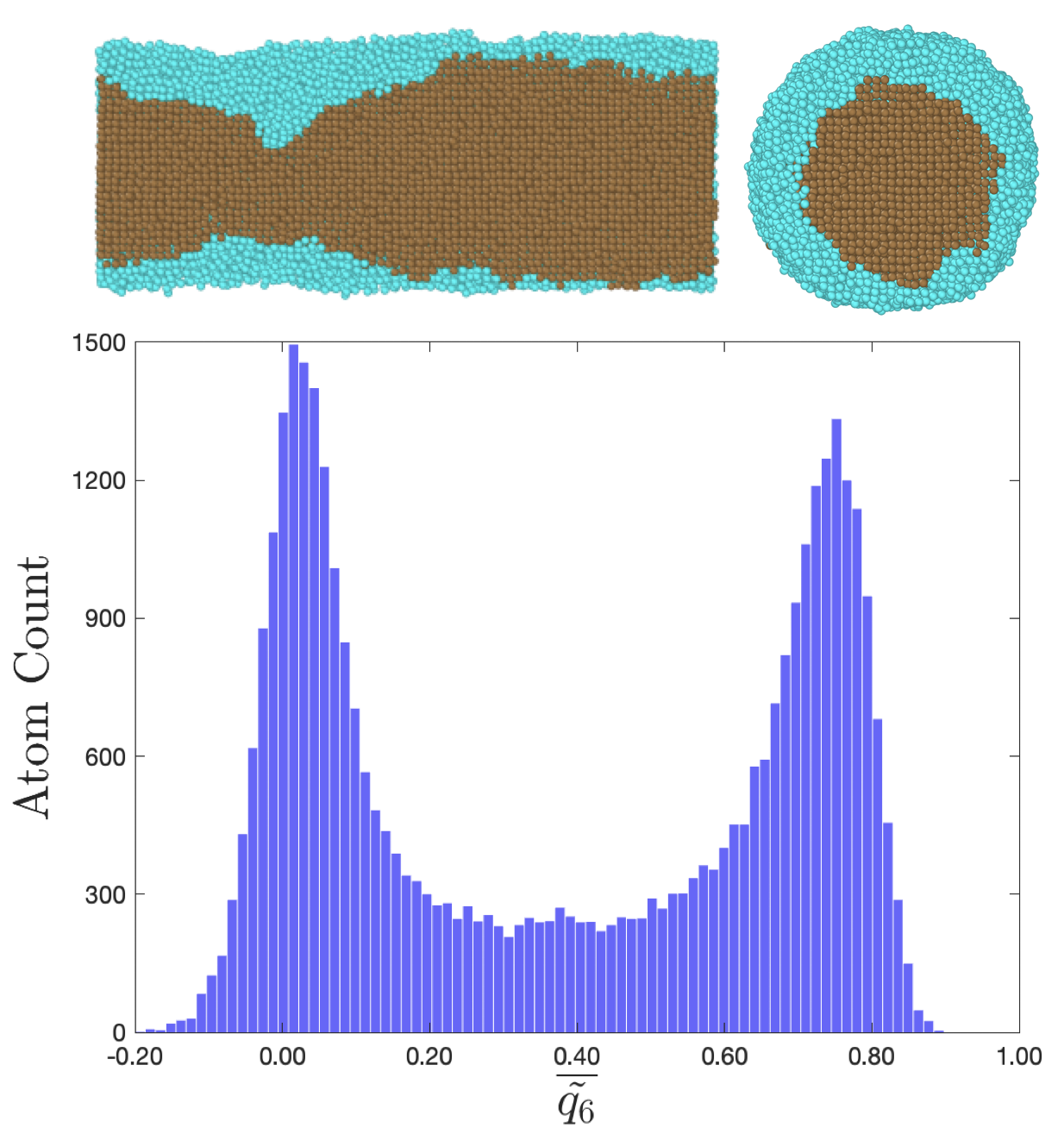}}
    \caption{Distribution of the order parameter $\overline{\tilde{q_6}}$ for a nanowire of initial radius $R_0=34$ \AA\ and length $L = 168$ \AA. Included are snapshots of the wire during this state looking (side on) down the \{100\} plane, and (top down) the \{001\} plane.}
	\label{Fig:q6bt_hist}
\end{figure}

Finally, we recover the order parameter seen in figure~\ref{Fig:q6bt_hist} by averaging the $\Tilde{q_{l}}(i)$ order parameters over their first coordination shell:
\begin{equation}
\label{qlm_5}
\bar{\tilde{q_{l}}}(i) =\frac{1}{1 + N_{\text{b}}(i)}\bigg(\tilde{q_{l}}(i) + \sum_{j \in N_{\text{b}}(i)} \tilde{q_{l}}(j)\bigg).
\end{equation}

Selecting $l=6$, we see figure~\ref{Fig:q6bt_hist}, where two distinct peaks emerge. One on the left indicating the presence of atoms in a `liquid-like' state, and one on the right, indicating atoms in a `solid-like' state. By taking a cutoff roughly in the middle of the peaks of the bimodal distribution we can defined which state every atom in the system should be (see \cite{ridings2019surface,ridings2022nanowire} for more details).
\vspace{0.5cm}
\\
In order the calculate the radius of the solid, values of $r(z)$ were binned along the wire axis and averaged, and then a nearest neighbour-type of algorithm is used to extract at `solid-like' atoms at the solid-liquid interface (see \cite{ridings2022nanowire} supplementary methods for more details).
\section{Molecular Dynamics Results}
\begin{figure}[htp]
	\centering
{\includegraphics[width=0.50\textwidth]
{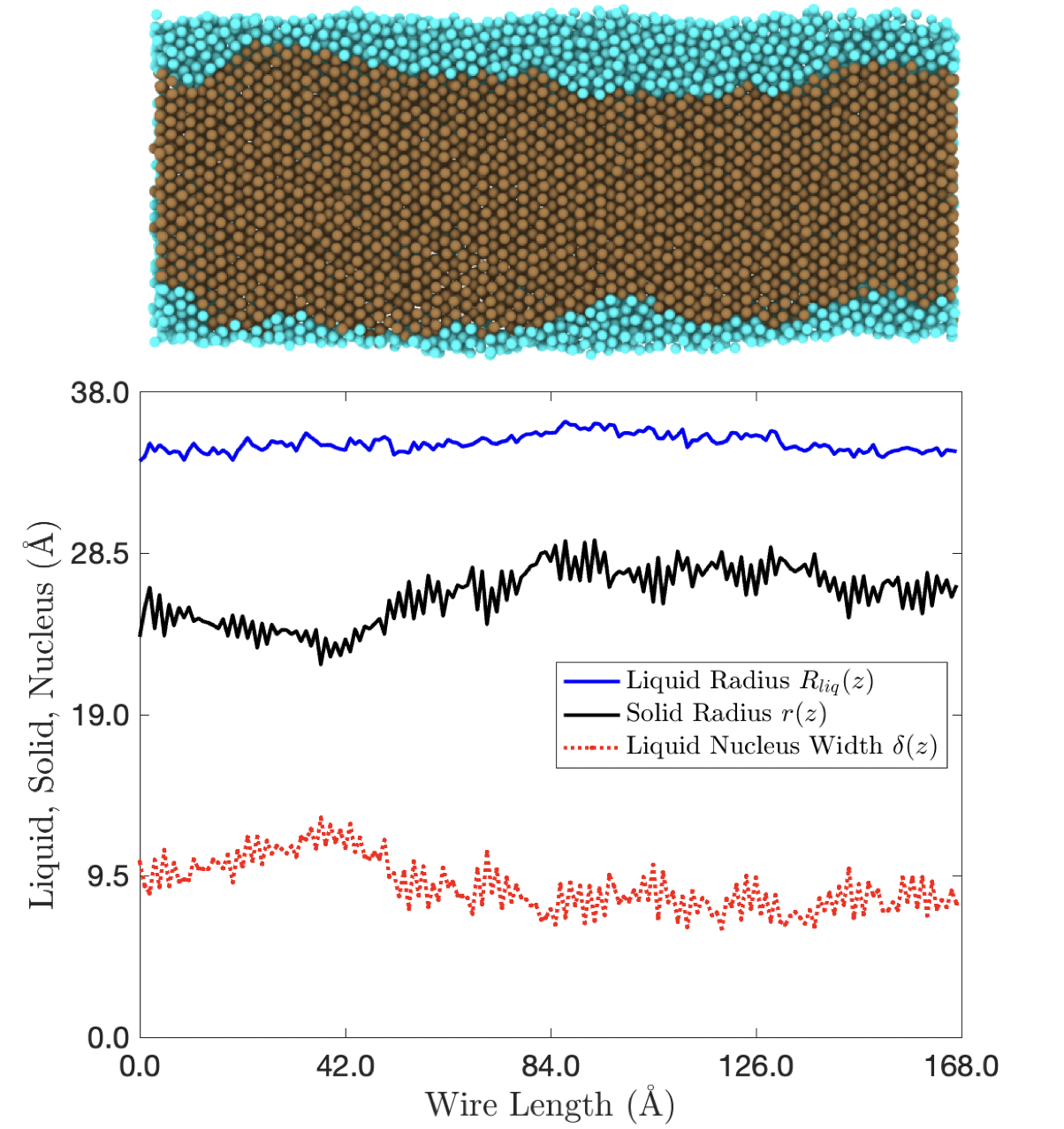}}
    \caption{How the liquid radius $R_\text{liq}(z)$, solid radius $r(z)$, and nucleus width $\delta(z)$ vary across the nanowire length. An accompanying snapshot is included where the values for each of the aforementioned are extracted from.}	
    \label{Fig:NW_Defns_MD_L40}
\end{figure}

\begin{figure}[htp]
	\centering
{\includegraphics[width=0.50\textwidth]
{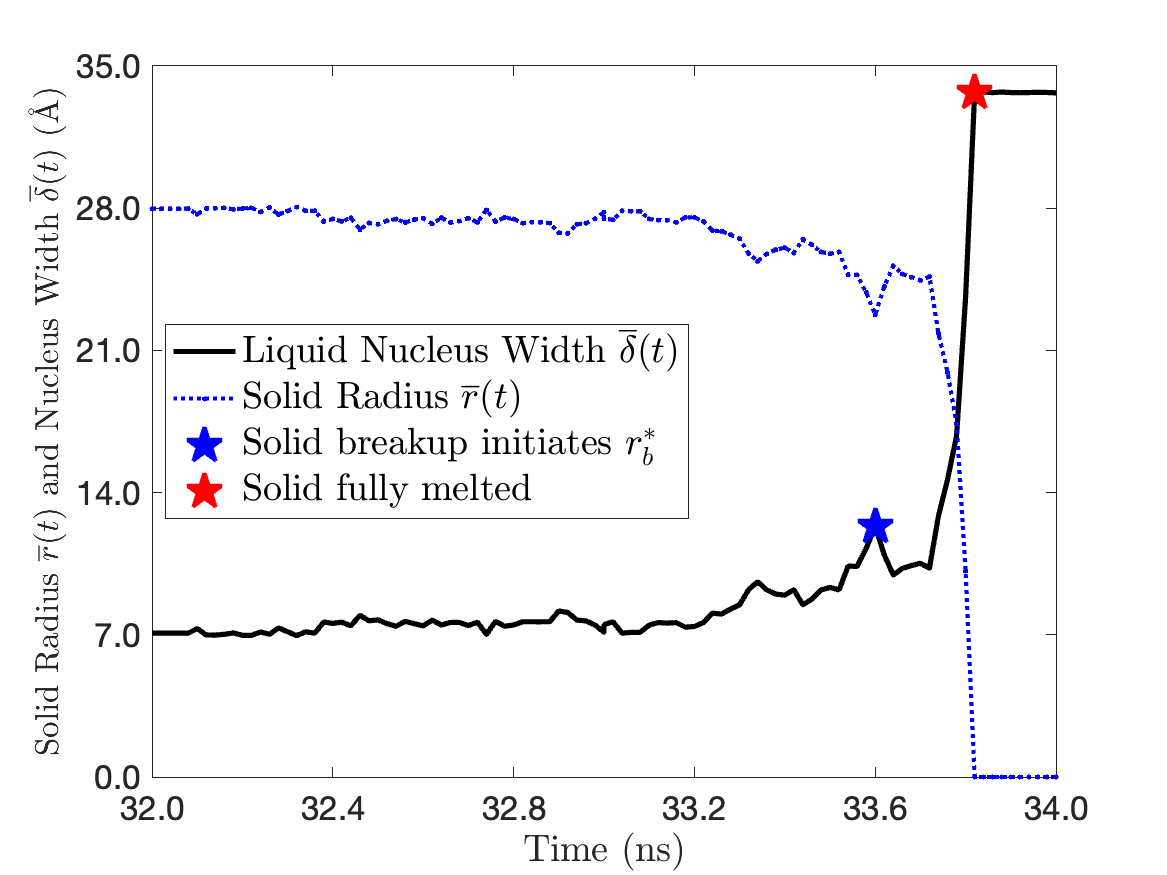}}
\caption{For the nanowire presented in figure~\ref{Fig:NW_Defns_MD_L40} we see the time evolution of $\overline{r(t)}$ and $\overline{\delta (t)}$ during the melting transition, where each point is averaged across the entire wire axis. Blue and red stars are included, indicating where the wire breakup $r^*_{\text{b}}$ initiates and where the wire is fully melted respectively.}
    \label{Fig:rstar_dstar}
\end{figure}

We begin by examining the results obtained from MD simulation where we extract the necessary data to estimate the interface velocity during the solid-liquid phase transition. In figure~\ref{Fig:NW_Defns_MD_L40} we see a plot of the definitions made in figure~\ref{Fig:NW_Defns}, namely $R_\text{liq}(z)$, $r(z)$ and $\delta(z)$, being the outside liquid radius, the inside solid radius, and the liquid nucleus thickness respectively. Accompanying the plot is a snapshot of the corresponding nanowire, where blue (light) circles represent the liquid, and brown (dark) circles represent the solid.
To find their time evolution, the average of $r(z)$, and $\delta(z)$ in figure~\ref{Fig:NW_Defns_MD_L40} are taken at each point in time during the solid-liquid phase transition. Looking to figure~\ref{Fig:rstar_dstar}, we can estimate the interface velocity by observing when the solid begins to pinch off, and when it has been fully consumed by the melt. By taking the gradient of $\overline{\delta}$ at the point where the solid pinches off (the blue star in figure~\ref{Fig:rstar_dstar}), and when the solid has completely melted (the red star in figure~\ref{Fig:rstar_dstar}), we get an estimate of $V(T)$ at $T=T_{\text{m}}$, which we can then compare to equation~\ref{Eq:int_velocity}. This gives an estimate of $d \delta / dt$, the rate at which the liquid nucleus consumes the solid as it melts. So, we assume the following relationship should hold 
\begin{equation}
    \label{eq:int_velocity_delta}
    \frac{d\delta}{dt} =C \,\exp(-Q / k_{\text{B}}T)(1-\exp(-\Delta \mu / k_{\text{B}}T)).
\end{equation}

\begin{figure}[htp]
	\centering
{\includegraphics[width=0.50\textwidth]
{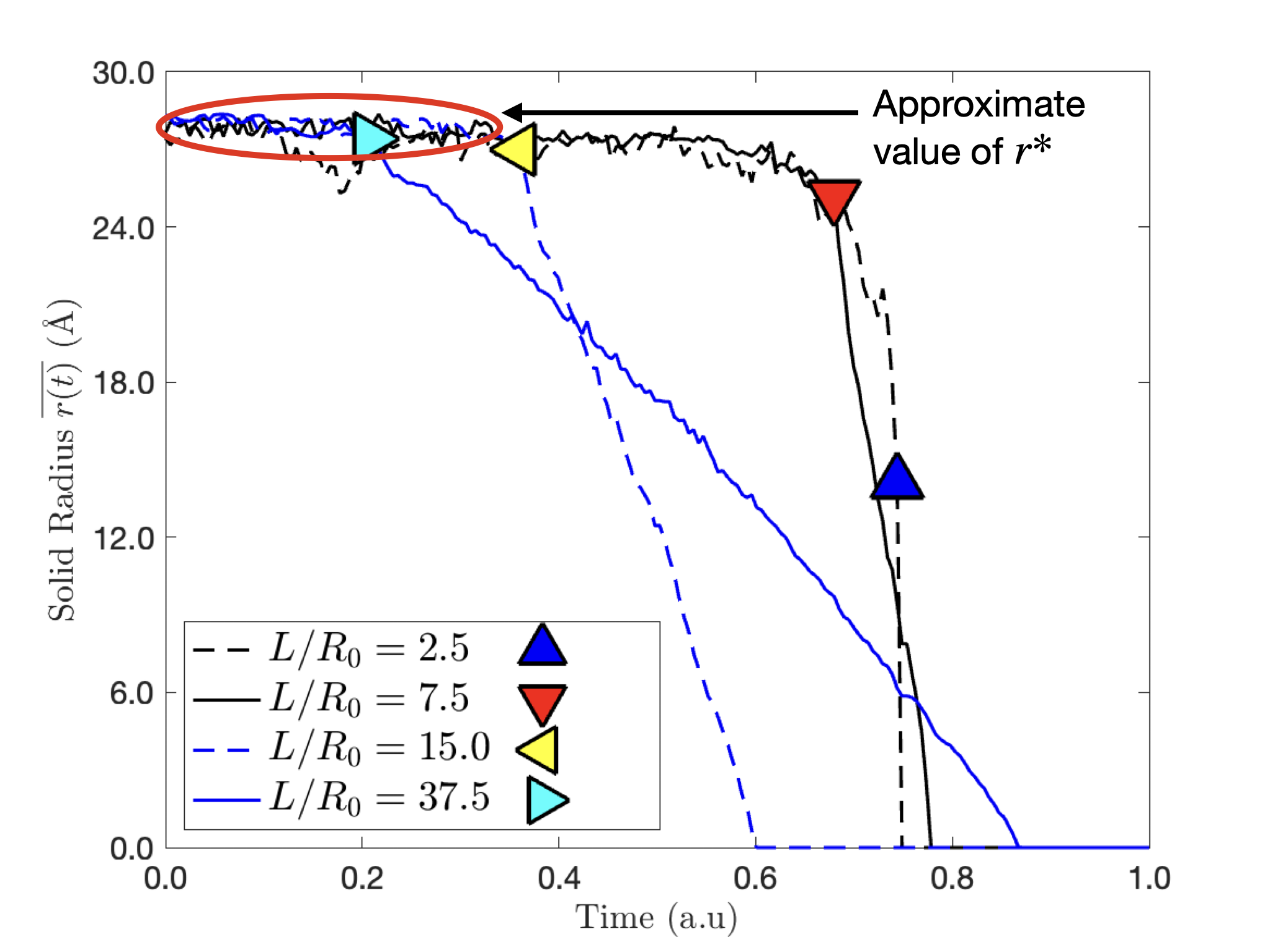}}
\caption{Here we see a figure of $\overline{r(t)}$ vs time for several wire aspect ratios. Prior to melting, the value of $\overline{r(t)}$ fluctuates about the same average. This is how we defined our values of $r^*$ from MD simulation. Each plot of $\overline{r(t)}$ here is approximately 2.0 ns, but have been scaled between 0 and 1 for clarity. Each line is accompanied by a marker showing the radius of the solid at the point where the solid pinches off, as indicated in the legend.}
    \label{Fig:rstar_time_LR0}
\end{figure}

\begin{figure}[htp]
	\centering
{\includegraphics[width=0.50\textwidth]
{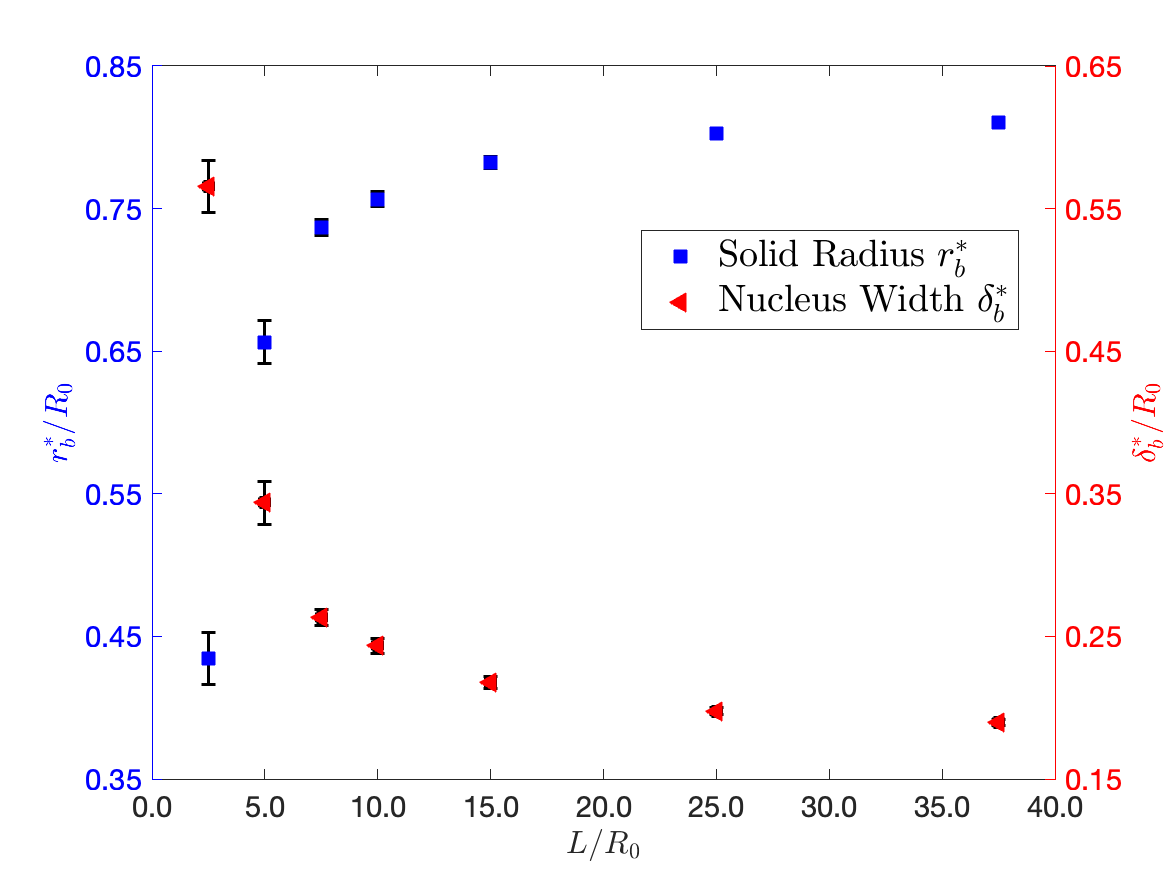}}
\caption{Here we see a figure of how the solid radius at the breakup, $r^*_{\text{b}}/R_0$ (blue axis), and the liquid nucleus thickness $\delta^*_{\text{b}}/R_0$ (red axis) changes for wire aspect ratio. As the wire lengths increase, these values tend to plateau off at approximately the value of $r^*$ defined by equation~\ref{eq:rstar}. Note that the error bars in longer wires are much smaller than in shorter wires, so their data point obstructs their value.}
    \label{Fig:rb_db_star}
\end{figure}

\begin{figure}[htp]
	\centering
{\includegraphics[width=0.50\textwidth]
{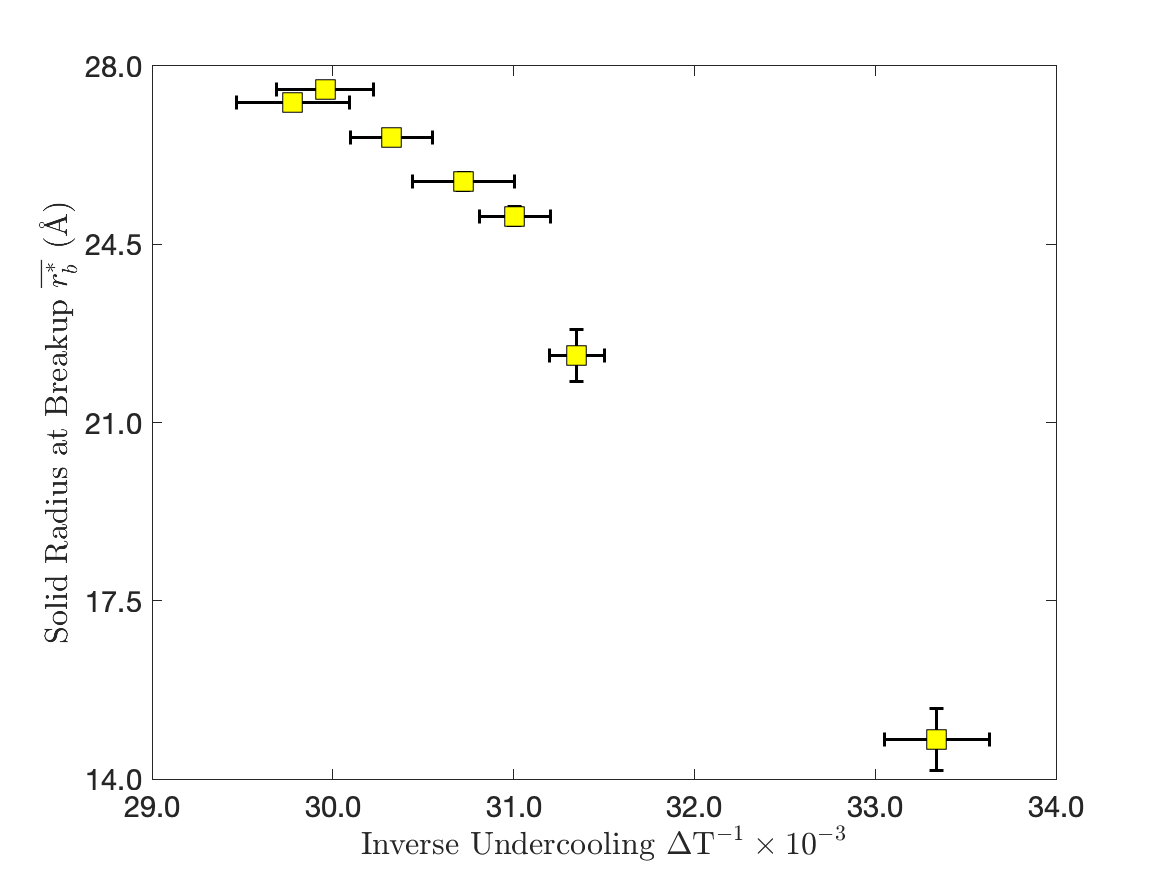}}
\caption{This figure shows how the solid radius at breakup $r^*_{\text{b}}$ runs proportional to $\Delta T^{-1}$, implied by equation~\ref{eq:rstar}.}
    \label{Fig:rbstar_DT}
\end{figure}

We then evaluate the result at $T=T_{\text{m}}$. To begin calculating $d \delta/dt$, we first estimate the value of $r^*$ seen in equation~\ref{eq:rstar} from MD simulation, and find the radius of the solid right before the breakup, $r^*_{\textrm{b}}$. To do this, we examine figure~\ref{Fig:rstar_time_LR0}, showing plots of $\overline{r(t)}$ versus time for different aspect ratios during the melting transition. Indicated by a coloured marked are the points where the solid begins to pinch off for each aspect ratio. The value of $\overline{r(t)}$ is approximately the same for each wire length as the melting point is approached. We take $r^*$ (as defined in equation~\ref{eq:rstar}) to be approximately 27.5 \AA\ for each simulation and wire length. With an initial radius $R_0=34$ \AA\ and $r^*=27.5$ \AA, given the lattice constant is $a=4.246$ \AA\ this shows us that about approximately 1.5 crystal layers melt prior to the initiation of bulk melting.
\vspace{0.5cm}
\\
In figure~\ref{Fig:rb_db_star}, we see $r^*_{\text{b}}/R_0$ and $\delta^*_{\text{b}}/R_0$, the solid radius and liquid nucleus width when the solid breaks up respectively, for increasing wire aspect ratio. These values differs for each wire aspect ratio and appears to plateau when the radius of the solid equals about $0.8R_0$, the same trend observed in previous work for copper nanowires \cite{ridings2022nanowire}. 
Previous work showed that this was due to the growth of instabilities that destabilise the interface prior to reaching the bulk melting temperature of the wire. Non-equilibrium effects that are taking place during the melting transition that allow the solid-liquid interface in shorter wires to remain stable down to smaller radii \cite{ridings2019surface,ridings2022nanowire}. 
\vspace{0.5cm}
\\
In figure~\ref{Fig:rbstar_DT} we look at how $r^*_{\text{b}}$ changes with undercooling, $\Delta T = T_{\text{c}} - T_{\text{m}}$. $T_{\text{m}}$ is calculated by taking the peak in the heat capacity defined by $C_{\text{v}} = dE/dT$, where $r^*_{\text{b}} \propto \Delta T^{-1}$ as expected. This implies that longer wires (with a higher $r^*_{\text{b}}$) tend to melt at a slightly lower point than their shorter counterparts, contrary to typically accepted theory where the melting point should depend only upon the initial radius $R_0$ \cite{kofman1994surface,gulseren1995premelting,ridings2019surface,ridings2022nanowire}.
\begin{figure}[htp]
	\centering
{\includegraphics[width=0.50\textwidth]
{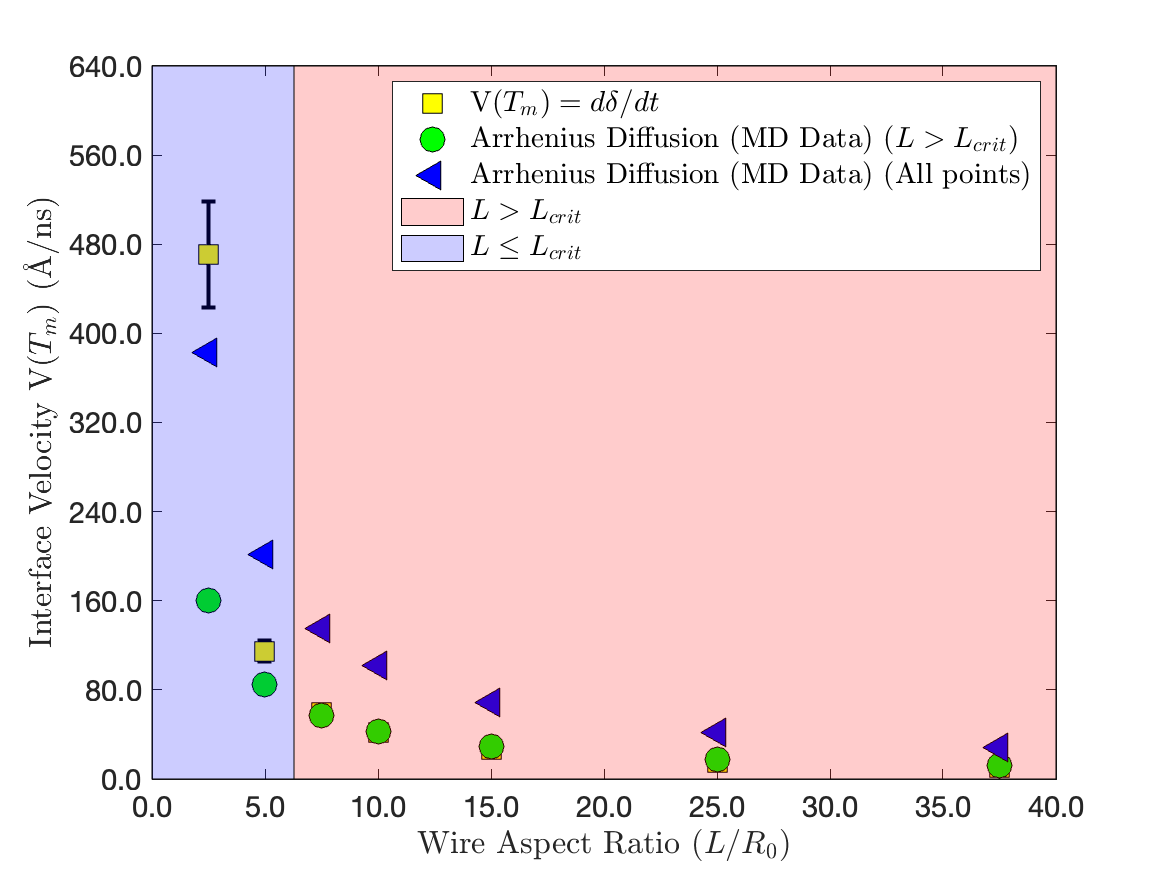}}
\caption{Here we see the averaged interface velocity of our sample wires plotted against the wire aspect ratio. The yellow squares represent the method where the gradient between the solid and liquid phases calculates the velocity. In contrast, the green circles use fits to equations~\ref{eq:Vel_Diff_Arrhenius} for wires with $L>L_{\textrm{crit}}$, and blue triangles use the same equation but fits to all data points. The light (pink) shaded region shows wires below the length $L\leq L_{\textrm{crit}}$, and the dark (blue) shaded region shows wires with $L> L_{\textrm{crit}}$.}
    \label{Fig:interface_velocity_1}
\end{figure}
\vspace{0.5cm}
\\
Now we evaluate the evolution of the solid-liquid interface velocity during the melting transition. We first find $d \delta/ dt$, the left-hand side of equation~\ref{eq:int_velocity_delta}, using the method outlined in figure~\ref{Fig:rstar_dstar}. Observing figure~\ref{Fig:interface_velocity_1}, we see the interface velocity follows a power law type relationship with the wire aspect ratio. Indeed, an Arrhenius-type diffusion model can fit the MD data (yellow squares) rather well, especially when only accounting for wires with $L>L_{\textrm{crit}}$ (green circles). When we fit all data points (blue triangles), the model captures the general behaviour, but offers a poorer fit. For wires with $L\leq L_{\textrm{crit}}$ we see a sharp increase in the interface velocity, telling us the timescale over which the melting occurs is quick. 
\vspace{0.5cm}
\\
Next we estimate the kinetic coefficient, $k(T)$. Different models suggest that the kinetic coefficient $k(T)$ indicates the mechanisms that dictate how the interface propagates \cite{sun2018mechanism}. Looking at equation~\ref{Eq:int_velocity}, we have MD data for $V(T)$ and $\Delta \mu / k_{\text{B}} T$ at $T=T_{\textrm{m}}$, meaning we can estimate $k(T)$ as
\begin{equation}
    \label{eq:kin_coeff_v_over_exp}
    k(T) = V(T)/(1-\text{exp}(-\Delta \mu / k_{\text{B}}T)).
\end{equation}

Using MD data at $T=T_{\textrm{m}}$ and the kinetic coefficient defined by equation~\ref{eq:Diff_Arrhenius}, all we need to do is to define $C$. Previous studies that assume diffusive behaviour define this constant $C$ as
\begin{equation}
    \label{eq:kin_coeff_constant}
    C = \frac{L_{\textrm{crit}}}{L}\frac{6 l D_0}{\lambda^2}.
\end{equation}

Here, we include a factor of $L_{\textrm{crit}} /L$ to describe the scaling behaviour (where $L_{\textrm{crit}}=2\pi R_0$), $l$ is a constant that represents a fraction of the lattice spacing, $\lambda$ is the mean-free path \cite{mazhukin2020atomistic}, and $D_0$ is the bulk diffusion constant at the melting temperature \cite{sun2018mechanism}. Thus, equation~\ref{eq:Diff_Arrhenius} becomes
\begin{equation}
    \label{eq:kin_coeff_model}
    k(T)= \frac{L_{\textrm{crit}}}{L}\frac{6 l D_0}{\lambda^2}\textrm{exp}\big(-Q/k_{\textrm{B}}T\big).
\end{equation}

Values of $l$, $\lambda$ and $Q$ from fits to wires with $L>L_{\textrm{crit}}$ and all wires are presented in table~\ref{tab:model_fit_params}.
\vspace{0.5cm}
\\
\begin{table*}[htb]
\caption{\label{tab:thermo_lit_vals}Bulk values/estimates of thermodynamic constants used in this study.}
\begin{tabular*}{\textwidth}{@{}l*{3}{@{\extracolsep{0pt plus
12pt}}l}}
\br
$T_{\text{c}}$ (K) & $\Delta h_{\text{m}}$ (eV/particle) & $L_{\text{v}}$ (eV/\AA$^3$) & $D_0$ (\AA$^2$/ns) \\
\mr
1255 \cite{sheng2011highly} & 0.118\cite{sun2018mechanism}  & 0.00610 \cite{lide2004crc} & 269.0\cite{sun2018mechanism} \\
\br
\end{tabular*}
\end{table*}
\begin{figure}[htp]
	\centering
{\includegraphics[width=0.50\textwidth]
{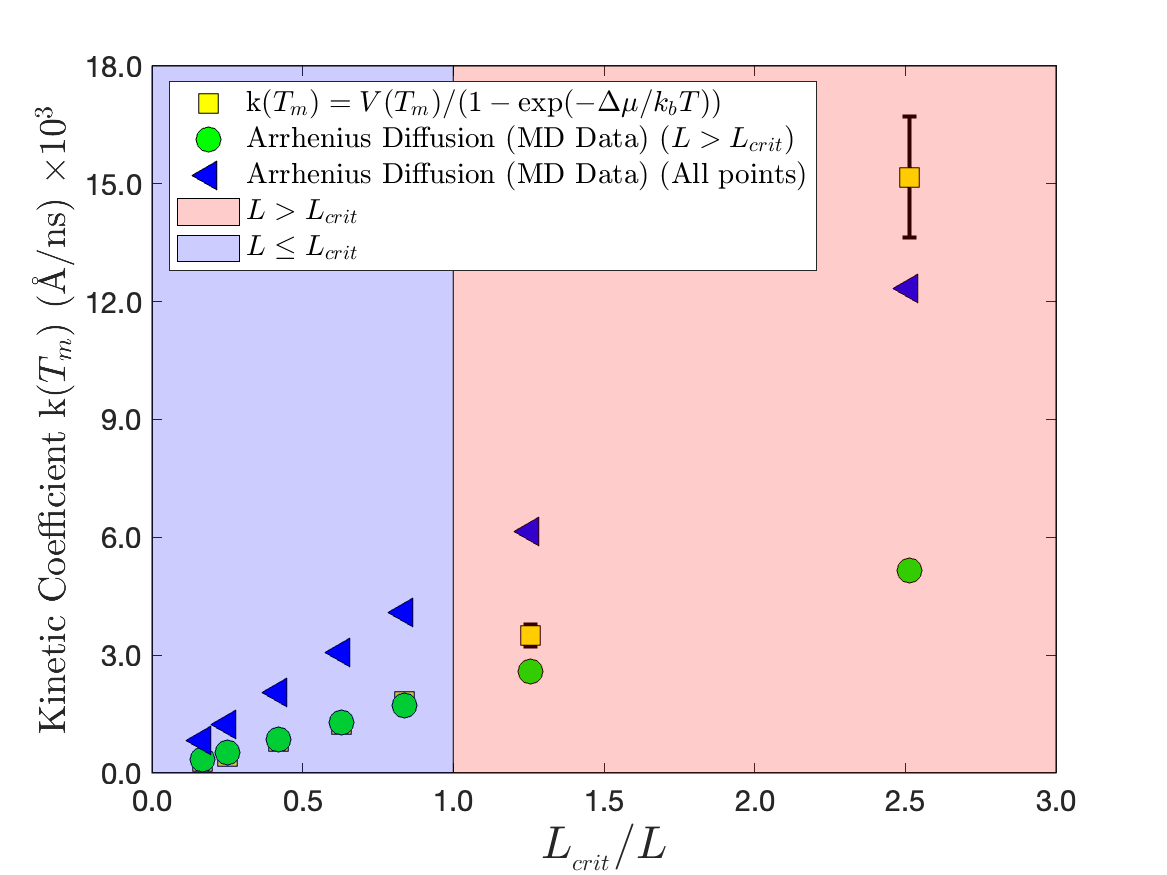}}
\caption{Values of the kinetic coefficient from MD data using equation~\ref{eq:kin_coeff_v_over_exp} (yellow squares), and fits to MD data using~\ref{eq:kin_coeff_model} for $L>L_{\textrm{crit}}$ (green circles) and all data points (blue triangles).}
    \label{Fig:kin_coeff_1}
\end{figure}
Figure~\ref{Fig:kin_coeff_1} shows us how the scaling behaviour directly influences the kinetic coefficient. We see that as the wire lengths increase, both fits appear to converge to the same values. For shorter wires with $L\leq L_{\textrm{crit}}$, the scaling behaviour becomes less dominant, where MD data (yellow squares) diverge away from the diffusion-type model. The results we have obtained from MD simulation and fits using equations~\ref{eq:int_velocity_delta} and~\ref{eq:kin_coeff_model} are summarised in table~\ref{tab:Int_vel_kinetic}.
\begin{table*}[htp]
\caption{\label{tab:model_fit_params}Values of the fitting parameters used in the Diffusion-type model. $l_{_{\text{long}}}$, and $\lambda_{_{\text{long}}}$ represent the $l$ and $\lambda$ from equation~\ref{eq:kin_coeff_constant} when fitting wires with $L>L_{\textrm{crit}}$, and $l_{_{\text{All}}}$, and $\lambda_{_{\text{All}}}$ represent when fitting to all wire lengths.}
\begin{tabular*}{\textwidth}{@{}l*{5}{@{\extracolsep{0pt plus
12pt}}l}}
\br
$l_{_{\text{long}}}$ (\AA) & $\lambda_{_{\text{long}}}$ (\AA) & $Q_{_{\text{long}}}$ (eV) & $l_{_{\text{All}}}$  (\AA) & $\lambda_{_{\text{All}}}$ (\AA) & $Q_{_{\text{All}}}$ (eV) \\
\mr
3.148 &0.703& 0.169& 4.575  & 0.397 & 0.238\\
\br
\end{tabular*}
\end{table*}
Comparing the velocities and kinetic coefficients calculated from MD and fits to using equations~\ref{eq:int_velocity_delta} and~\ref{eq:kin_coeff_model}, we see good agreement for $L >L_{\text{crit}}$, but poor results when fitting to all wire lengths. This tells us that for wires $L<L_{\text{crit}}$ a simple Arrhenius-type diffusion model is adequate enough to describe the observed melting mechanism. The scaling term $L_{\text{crit}}/L$ has a strong presence for longer wires, and matters less below the critical length. This shows a diffusion-type model can aptly describe the mechanisms dictating the melting behaviour for wires $L>L_{\text{crit}}$, but we still require an explanation that accounts for the difference in shorter wires.
\vspace{0.5cm}
\\
\begin{table*}[htp]
\caption{\label{tab:Int_vel_kinetic}A table showing the magnitudes of the interface velocities and kinetic coefficients calculated from MD, fitting to wires $L\geq L_{\text{crit}}$, fitting to all wires respectively. Kinetic coefficients are scaled by $\times 10 ^3$. Besides aspect ratio, $L/R_0$, all units are in \AA/ns.}
\begin{tabular*}{\textwidth}{@{}l*{6}{@{\extracolsep{0pt plus
12pt}}l}}
\br                              
$L/R_0$ & $V_{_\text{MD}}$ & $V_{_{\text{long}}}$ & $V_{_\text{All}}$ & $k_{_\text{MD}}$ & $k_{_{\text{long}}}$ & $k_{_\text{All}}$ \\
\mr
2.5 & 471.1 & 160.4 & 382.9 & 15.2 & 5.16 & 12.3  \\ 
5.0 & 115.0 & 84.4 & 201.3 & 3.50 & 2.58 & 6.15  \\ 
7.5 & 60.0 & 56.8 & 135.4 & 1.81 & 1.72 & 4.09  \\ 
10.0 & 41.4 & 42.9 & 102.3 & 1.24 & 1.28 & 3.07  \\ 
15.0 & 26.7 & 28.9 & 68.9 & 0.791 & 0.858 & 2.04  \\ 
25.0 & 14.7 & 17.6 & 42.0 & 0.429 & 0.514 & 1.22  \\ 
37.5 & 10.3 & 11.7 & 27.9 & 0.301 & 0.343 & 0.817  \\ 
\br
\end{tabular*}
\end{table*}
\begin{table*}[htp]
\caption{\label{tab:thermo_quantities}A table showing quantities calculated in this study. All quantities are assumed to be calculated for $T=T_{\text{m}}$.}
\begin{tabular*}{\textwidth}{@{}l*{4}{@{\extracolsep{0pt plus
12pt}}l}}
\br                              
$L/R_0$ & $T_{\text{m}}$ (K) & $r^*_{\text{b}}$ (\AA) & $\delta^*_{\text{b}}$ (\AA) & $\tau^*$ (ns) \\
\mr
2.5 & 1220.0 & 14.7 & 19.2 & 0.033\\ 
5.0 &  1218.1 & 22.3 & 11.7 & 0.19\\ 
7.5 & 1217.7 & 25.05 & 8.94 & 0.39\\ 
10.0 &  1217.5 & 25.7 & 8.27 & 0.59\\ 
15.0 & 1217.0 & 26.6 & 7.40  & 0.94\\ 
25.0 &  1216.4 & 27.3 & 6.71 & 1.76\\ 
37.5 & 1216.6 & 27.5 & 6.45 & 2.58\\ 
\br
\end{tabular*}
\end{table*}
\vspace{0.5cm}
\\
Experimental results of laser melting Au foils differentiates between homogeneous (short time-scale) and heterogeneous (long time-scale) where homogeneous or heterogeneous melting dictate the pathway a solid takes during the melting transition \cite{mo2018heterogeneous}. Computational studies also imply that the difference in time-scales of the melting phenomena indicate whether or not a solid has undergone homogeneous/heterogeneous nucleation/melting, where homogeneous melting occurs more rapidly, with the solid being heated above its stability limit \cite{ivanov2003combined,lin2006time}. The relationship between the melting time and temperature $T_{\text{m}}$ indeed shows that wires whose solid is rapidly consumed melt at slightly elevated temperatures. To explore the mechanisms dictating the melting behaviour below the critical length, we looked at the time-scale the melting takes place over for the wires studied. 
\begin{figure}[htp]
	\centering
{\includegraphics[width=0.50\textwidth]
{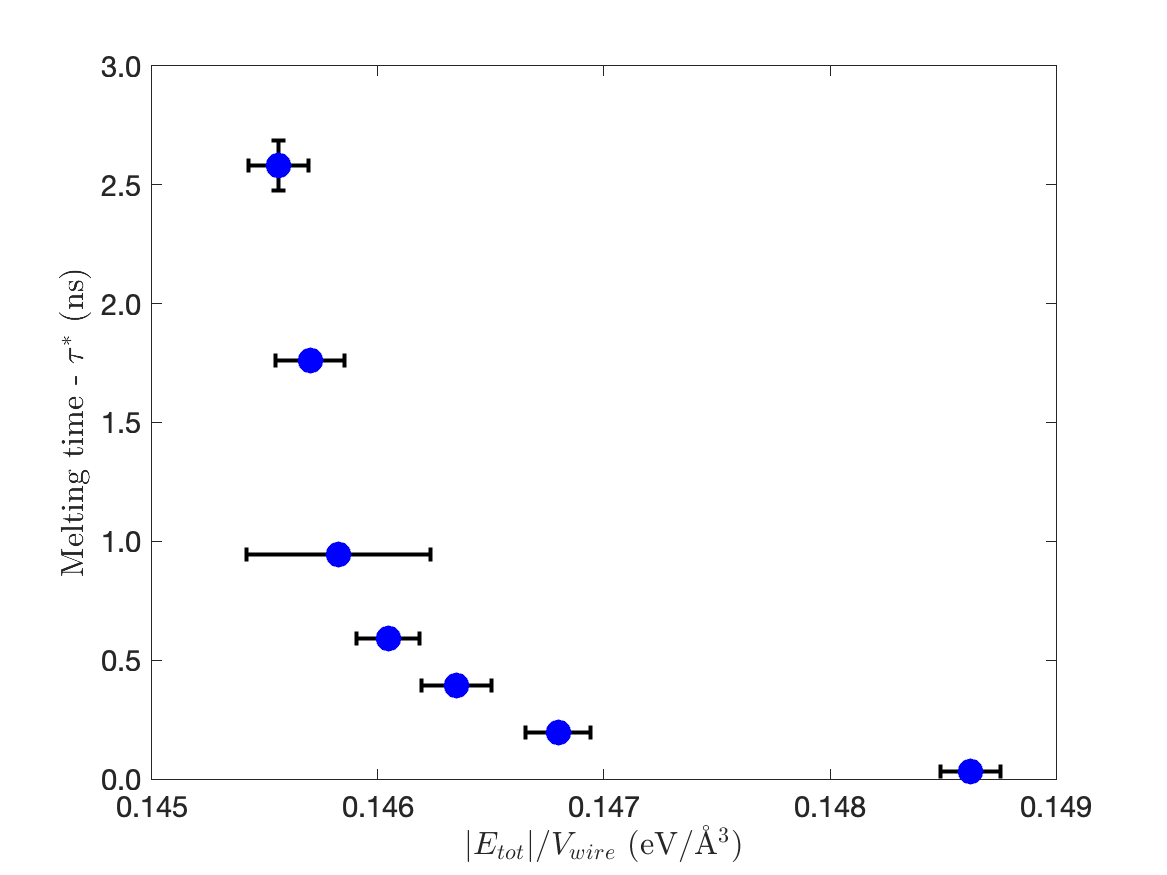}}
\caption{The time-scale of the melting transition changing with energy density for each nanowire. The longest nanowires take 1 - 3 ns to be consumed, whereas the shortest ones melt rapidly.}
    \label{Fig:Eden_tmelt}
\end{figure}
First, we define the melting time $\tau^*$ as the time it takes from just before the breakup $t_{\text{b}}$ to when the solid has fully melted $t_{\text{m}}$ (i.e. the time-scale we use to define the interface velocity in figure~\ref{Fig:rstar_dstar}). Next, we define the energy density by taking the absolute value of the total energy at the melting point $T_{\text{m}}$, and dividing it by the total volume of each wire. This is shown in figure~\ref{Fig:Eden_tmelt} where we see the timescales over which the solid is consumed during the melting transition. The longest nanowire takes about 2.58 ns (2580 ps) to from breakup to fully liquid, versus the shortest which takes 0.033 ns (33 ps). 
\begin{figure}[htp]
	\centering
{\includegraphics[width=0.50\textwidth]
{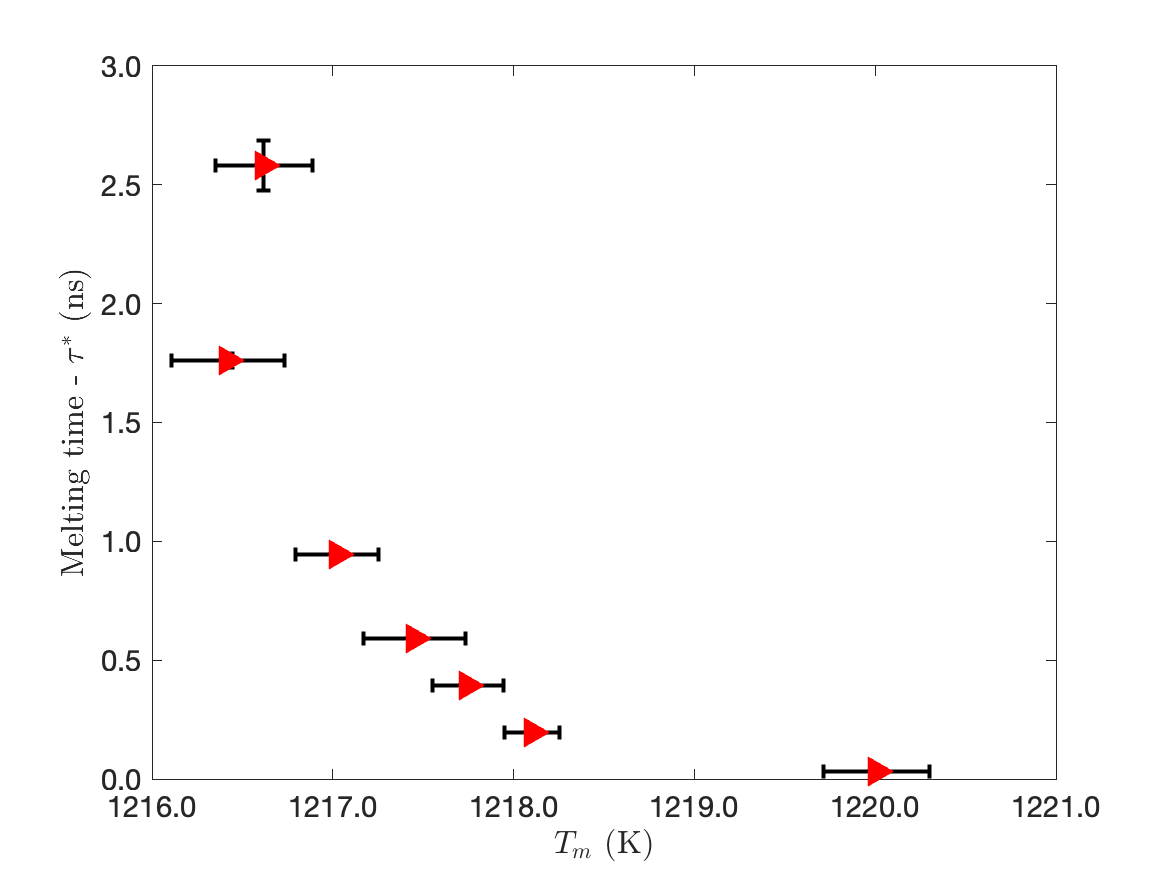}}
\caption{The time-scale of the melting transition changing with energy density for each nanowire. The longest nanowires take 1 - 3 ns to be consumed, whereas the shortest ones melt in tens of picoseconds.}
    \label{Fig:tmelt_Tmelt}
\end{figure}
Figure~\ref{Fig:tmelt_Tmelt} shows that the shortest wires all melt quicker. These results agree with the sentiments from a previous study, suggesting that longer wires will melt at a slightly lower temperature than their shorter counterparts \cite{ridings2022nanowire}, where in the case of the shorter wires, it appears that they can be heated slightly above the stability limit of their crystal structure, meaning that the liquid nucleus will prefer to grow rather than break apart the solid like it does with longer wires. 
\section{Discussion}
By constructing a interface velocity model assuming Arrhenius diffusion and utilizing thermodynamic quantities estimated from MD simulation, we describe the mechanisms that dictate the observed pathways the solid takes in simulations of melting in metal nanowires. In wires $L>L_{\text{crit}}$, the solid will pinch off at a point along the wire axis, and melt along its length until the solid has been fully consumed. For longer wires, there is good agreement between the data obtained from MD simulation, and predictions made by a simple diffusion-based type of model, and as wires become shorter, the model deviates away from the observed data. This implies that as wire lengths becomes shorter than the critical length, non-equilibrium effects become more pronounced and the general models used to describe interface velocities break down. For wires $L\leq L_{\text{crit}}$ there is a preference of the liquid nucleus at the solid-liquid interface to move from the outside towards the centre, consuming the solid in its path. These wires deviate away from predictions made by the constructed interface velocity model, with the scaling behaviour unable to account for the rapid consumption of the solid. For these shorter wires, the observed melting mechanism appears to be dictated by a tendency of the interface to be stabilised with respect to the melt, allowing for the inside solid to become slightly overheated, leading to a much more rapid melting time. The heating rate is identical in all wires, so the amount of energy given to the system at each timestep in order to maintain a constant temperature is the same with the only difference being the volumes of each sample. This agrees with the observed experimental results that demonstrate higher energy densities lead to a rapidly consumed solid \cite{mo2018heterogeneous}. 
Classical nucleation theory predicts that the energy barrier to melting is inversely proportional to the undercooling, that is, $\Delta G^* \propto \Delta T^{-1}$. Indeed, for longer wires, this fact has been observed, with longer wires consistently melting at a slightly lower temperatures than their shorter counter parts. This has been observed in simulations both in this, and previous work, where the solid becomes destabilised in longer wires as the melting temperature is approached. This difference in nucleation pathways has been noted to result in much faster melting times, as can be seen for the results from this study. The average time it takes for the shortest nanowires to melt is on the order of several tens of picoseconds, as opposed to nanoseconds for longer wires. This could partly be due to the solid in the shorter wires slightly overheating past their `classically determined' reduced melting point for their given radius, meaning the solid can be more rapidly consumed as a result. Moreover, the crystal structure underlying the liquid nucleus has an influence on the melting kinetics \cite{fan2021localization}, \cite{montero2019interfacial, montero2020interfacial}. All this points to a slight difference in the energy barrier between the solid and liquid phases, dictating the mechanisms to initiate a phase transition. Researchers have cited that the separation of defect pairs is a common mechanism for reducing nucleation barriers \cite{mochizuki2013defect, cordin2014experimental}, where investigating the role of defects in the melting mechanisms and shift in energy barriers in metal nanowires would be the next logical step in this work.
\section{Conclusion}
This study elucidates the mechanisms driving the melting of silver nanowires, highlighting the critical role of nanowire geometry in determining melting pathways. An Arrhenius-type model successfully captures the melting behavior for wires longer than the critical length. However, for shorter wires, deviations arise due to non-equilibrium effects, including rapid overheating of the solid core, stabilization of the solid-liquid interface, and the influence of high energy densities on melting. These mechanisms result in faster phase transitions and distinct pathways compared to their longer counterparts. These insights enhance our understanding of nanoscale phase transitions and could inform the design of stable nanostructures for applications in nanotechnology and materials science. Future work should explore the role of defects, surface curvature, and facet-specific interfacial properties to refine models for the melting of shorter nanowires.
\section*{References}
\bibliographystyle{unsrt}
\bibliography{Ag_stability}
\end{document}